\documentclass[twocolumn]{aastex631}

\usepackage{color}
\usepackage{bm}
\usepackage{ amssymb, amsmath }

\newcommand{\grad}{\bm{\nabla}}
\newcommand{\g}{\textbf}

\newcommand{\YY}{\bm{Y} }

\newcommand{\bb}{\bm{b} }

\newcommand{\rr}{\bm{r} }

\newcommand{\vv}{\bm{v} }
\newcommand{\xx}{\bm{x} }

\newcommand{\zz}{\bm{z} }

\renewcommand{\bell}{\boldsymbol{\ell}}

\shorttitle{SHORT TITLE}
\shortauthors{Pecora et al.}
\graphicspath{{./}{./figs_tmp/}}

\begin{document}

\title{Three-dimensional energy transfer in space plasma turbulence from multipoint measurement}

\author[0000-0003-4168-590X]{Francesco Pecora}
\email{fpecora@udel.edu}
\affil{Department of Physics and Astronomy, University of Delaware, Newark, DE 19716, USA}

\author[0000-0001-8184-2151]{Sergio Servidio}
\affil{Universit\`a della Calabria, Arcavacata di Rende, 87036, IT}

\author[0000-0003-2965-7906]{Yan Yang}
\affil{Department of Physics and Astronomy, University of Delaware, Newark, DE 19716, USA}

\author[0000-0001-7224-6024]{William H. Matthaeus}
\affil{Department of Physics and Astronomy, University of Delaware, Newark, DE 19716, USA}

\author[0000-0001-8478-5797]{Alexandros Chasapis}
\affil{Laboratory for Atmospheric and Space Physics, University of Colorado Boulder, Boulder, CO, 80309, USA}

\author[0000-0001-5680-4487]{Antonella Greco}
\affil{Universit\`a della Calabria, Arcavacata di Rende, 87036, IT}

\author[0000-0003-1304-4769]{Daniel~J. Gershman}
\affiliation{NASA Goddard Space Flight Center, Greenbelt, Maryland 20771, USA}

\author[0000-0001-8054-825X]{Barbara~L. Giles}
\affiliation{NASA Goddard Space Flight Center, Greenbelt, Maryland 20771, USA}

\author[0000-0003-0452-8403]{James~L. Burch}
\affiliation{Southwest Research Institute, San Antonio, Texas 78238, USA}



\begin{abstract}
A novel multispacecraft technique applied to Magnetospheric Multiscale (MMS) mission data collected in the Earth's magnetosheath enables evaluation of the energy cascade rate solving the full Yaglom's equation in a turbulent space plasma. The method differs from existing approaches in that 
(i) it is inherently three-dimensional;
(ii) it provides a statistically significant number of estimates from a single data stream; and
(iii) it allows for a direct visualization of energy flux in turbulent plasmas. This new technique will ultimately provide a realistic, comprehensive picture of the turbulence process in plasmas.
\end{abstract}

\keywords{Suggested keywords}


\section{Introduction}
\label{sec:intro}

A long-standing and central problem in turbulence is how energy is transferred across scales from large-scale reservoirs to small-scale dissipation. The problem relates to fluids and plasmas throughout the universe, but until recent decades was studied almost exclusively in the hydrodynamic context. Experimental study of energy transfer has always been crucial in the development of the subject \citep{MoninYaglom, Pope}but until now a full three-dimensional evaluation of the Yaglom cascade rate has not been available in turbulent plasmas. We present such an evaluation below, using high-resolution data from the Magnetosphere Multiscale (MMS) mission in the terrestrial magnetosheath. 

Initial efforts by \citet{karman1938statistical} provided monumental results that describe turbulence in terms of double and triple correlations of a fluctuating velocity field, indicating a balance between time dependence, nonlinear transfer, and dissipation. Based on the  von K\'arman-Howarth (vKH) equations, \citet{kolmogorov1941dissipation} derived an exact relation between the third-order structure function and the rate at which energy is ``cascaded'' through scales. A few years later, \citet{yaglom1949local} derived a similar expression for a passive scalar (temperature) in a turbulent flow, which differs from Kolmogorov's exact law in that the third-order structure function is mixed, involving both the velocity and the temperature. All the above results were derived for homogeneous isotropic incompressible hydrodynamic flows.

\citet{politano1998dynamical_PP1, Politano1998vonkarman_PP2} generalized the hydrodynamic results to  incompressible magnetohydrodynamics, a model often adopted as a good description of space plasma \citep{Barnes1979hydromagnetic}. They extend the vKH equation to include the properties of a magnetized fluid described by the Els\"asser fields $\zz^\pm = \vv \pm \bb$, where $\vv$ is the velocity and $\bb$ the magnetic field in Alfv\'en units. The vKH equation in MHD reads:
\begin{equation}
    \displaystyle\frac{\partial}{\partial t}\langle |\delta \mathbf{z}^\pm|^2 \rangle = - \mathbf{\nabla}_{\boldsymbol{\ell}} \cdot \langle \delta \mathbf{z}^\mp |\delta \mathbf{z}^\pm|^2 \rangle + 2 \nu \nabla^2_{\boldsymbol{\ell}}\langle |\delta \mathbf{z}^\pm|^2 \rangle - 4\epsilon^\pm
    \label{eq:vKH}
\end{equation}
where $\delta\zz^\pm(\bell) = \zz^\pm(\xx) - \zz^\pm(\xx + \bell)$ are increments of the Els\"asser fields at a certain vector lag $\bell$, derivatives $\mathbf{\nabla}_{\boldsymbol{\ell}}$ are taken in the lag (increment) space and $\epsilon^\pm$ is the energy dissipation (or cascade) rate. 
The inertia term on the left side of Eq.\ref{eq:vKH} usually dominates at very large scales, the nonlinear term peaks at separations that fall into the inertial range, while the dissipative term becomes important at very small scales. The sum of these three terms gives (four times) the mean dissipation rate.
In a system that is sufficiently well-separated in scales, each of the different terms of Eq.\ref{eq:vKH} becomes dominant in magnitude in a range of scales in which the other terms are negligible. This condition is usually associated with high Reynolds numbers or a very large system.  When the system has a lower Reynolds number (e.g. in simulations), this separation of scales is not pristine, and different contributions ``leak'' into neighboring scale ranges. In this case, the evaluation of each term will give a partial cascade rate estimate as it will be missing the contribution from the other terms that became relevant at those scales.

When applying this theory to space data, the time derivative term is usually inaccessible given that the spacecraft does not follow the plasma flow \citep{taylor1938spectrum_frozenin}. The dissipative term on the other hand requires the knowledge of or assumptions about the viscosity, which itself is a separate problem, especially for plasma \citep{pezzi2021dissipation}. Usually one focuses on the nonlinear term, assuming good scale separation. The remaining relationship is 
\begin{equation}
    \mathbf{\nabla}_{\boldsymbol{\ell}} \cdot \langle \delta \mathbf{z}^\mp |\delta \mathbf{z}^\pm|^2 \rangle = - 4\epsilon^\pm
    \label{eq:yaglom}
\end{equation}
that involves a mixed third-order structure function and, in analogy with \citet{yaglom1949local}, $\langle \delta \mathbf{z}^\mp |\delta \mathbf{z}^\pm|^2 \rangle = \YY$ is generally called Yaglom flux, and the above equation Yaglom's law $\grad_\ell \cdot \YY^\pm = -4\epsilon^\pm$. The total dissipation rate is $\epsilon = ( \epsilon^+ + \epsilon^-)/2$.
This equation can be applied to anisotropic systems since vKH equation is derived without isotropy assumption. When the system is not well-separated in scales or available lags are not in the inertial range, Eq.~\ref{eq:yaglom} provides a partial answer, and the result is interpreted as the contribution to the total energy cascade rate from the nonlinear term.

Currently, our capability of measuring (partial) cascade rates in space plasmas has been limited to directional averages \citep{osman2011directional,bandyopadhyay2018incompressive} or, assuming isotropy, the 1D version of Eq.~\ref{eq:yaglom} \citep{sorrisovalvo2007observation, macbride2008turbulent, carbone2009scaling, hadid2017energy, bandyopadhyay2020insitu, andres2022incompressible}. Recent numerical studies \citep{wang2022strategies, jiang2023energy} showed that the directional dependence of Yaglom's equation is not negligible and therefore isotropy is not a good assumption, and instead,
one requires measurements in several directions in 3D lag space.

Below we will describe a technique that enables the full divergence of Eq.~\ref{eq:yaglom} to be calculated in lag space and obtain several estimates through which a statistical evaluation of the cascade rate will be possible.



\section{Data}

We use data from MMS \citep{burch2016magnetospheric} for the magnetic field from the Fluxgate magnetometers \cite[FGM,][]{russell2016magnetospheric}, and ion density and velocity from the Fast Plasma Investigation \citep[FPI,][]{pollock2016fast}. Measurements are available in burst mode at cadences of 128~Hz and 150~ms respectively. For velocities, spintone has been removed. We use electron density, as it generally is more accurately determined, and assume quasi neutrality. Density signals have been checked not to have values larger than 50~cm$^{-3}$ that can be polluted by instrumental inaccuracies. All data have been resampled to the lowest common cadence of 150~ms. We analyzed several intervals from \citep{yang2023agirotropy} during which MMS was in the magnetosheath. Properties of these intervals are listed in Table~\ref{tab:table1}, including their measured mean turbulence cascade rates.
\begin{table*}[ht]
    \centering
    \begin{tabular}{ccccccccc}
        & Date (UTC) & $\delta b / B$ & $\delta \rho / \rho $ & $\beta$ & $\tau_c$(s) & $\lambda_c$(km) & $d_i$(km) & $\langle \epsilon \rangle$ ($10^6$~J kg$^{-1}$ s$^{-1}$)  \\
        \hline
        I   & 2017 Sep 28 06:31:33 -- 07:01:43 & 0.51 & 0.19 & 5.9 & 81.0 & 41229 & 46 & $-1.5 \pm 0.8$ \\
        II  & 2017 Nov 10 22:35:43 -- 22:52:03 & 3.17 & 0.43 & 8.3 & 2.9  & 1377 & 74 & $22.5 \pm 9.5$ \\
        III & 2017 Dec 21 07:21:54 -- 07:48:01 & 1.92 & 0.31 & 4.7 & 6.1  & 652 & 50 & $24.8 \pm 13.4$ \\
        IV  & 2017 Dec 26 06:12:43 -- 06:52:23 & 0.82 & 0.21 & 4.5 & 16.3 & 3712 & 48 & $1.6  \pm 0.4$ \\
        V   & 2018 Apr 19 05:10:23 -- 05:41:53 & 2.99 & 0.29 & 15.0 & 5.1 & 1168 & 36 & $1.9  \pm 0.7$
    \end{tabular}
    \caption{Reported are the dates of the analyzed intervals, magnetic and density fluctuation levels, plasma beta (ratio between kinetic and magnetic pressure), correlation time, ion inertial length, and the mean energy cascade rate with associated uncertainties calculated as the width of the respective histograms. }
    \label{tab:table1}
\end{table*}

\section{Technique}

We implemented the novel technique described in \citep{pecora2023helioswarm} through which it is possible to have several estimates of the cascade rate using Yaglom's law, Eq.~\ref{eq:yaglom}. The method also accommodates arbitrary rotational symmetries of the fluctuations. This technique was applied to Helioswarm-like trajectories in simulations, and the number of possible estimates was of the order of 58000. Here we will add another variation that affords a larger pool of estimates. This is particularly useful with MMS, as the number of estimates will not be as large. Nevertheless, the technique provides a statistically significant sample.

The technique is based on the fact that derivatives have to be computed in the lag space to evaluate Yaglom's equation. Indeed, it is well known that 4 spacecraft in real space are connected by 6 baselines. Since we need to compute increments such as $\delta\zz_{ij}(\rr_{ij})=\zz_i(\rr_i)-\zz_j(\rr_j)$, ${\boldsymbol{\ell}}=\rr_{ij}$, it means that we will have 6 possible values of $\delta\zz(\rr_{ij})$ in lag space where $i,j=1,\dots 4,~\;~i<j$ represent each possible pair of spacecraft, with spacecraft $i$ at position $\rr_i$. Here is the additional step of the technique: Since the choice of the order of the pair is arbitrary, we can have another 6 points in lag space by just reverting the pairs. It means that we will have a total of 12 points in lag space with $\delta\zz_{ji}=-\delta\zz_{ij}$. Of course, these additional points will not give independent information but will be valuable in the construction of tetrahedra as we will describe below.

Since we want to use algorithms that have been well-tested using MMS data in the past, we examine only tetrahedral configurations; that is, we rearrange our points in lag space in sets of 4. The number of non-repeating partitions of K elements out of N total elements is given by $\binom{N}{K}=\frac{N!}{K!(N-K)!}$. If we were to use the 6 points in lag space that come from $\delta\zz_{ij}(\rr_{ij})$, we would have $\binom{6}{4} = 15$ possible tetrahedra. When we add $\delta\zz_{ji}(\rr_{ji})$, our available points become 12, with a total of $\binom{12}{4} = 660$ tetrahedra. Evidently, not all configurations are independent, indeed half of them will give duplicated results leaving 330 independent estimates of Eq.~\ref{eq:yaglom}.

We will need to run one additional check for the quality of tetrahedra. Indeed, well-behaved real-space configurations do not necessarily translate into well-behaved lag-space tetrahedra. For the quality check of the 330 tetrahedra in lag space, we make use of the elongation-planarity parameters described in \citet{paschmann1998analysis} and we briefly recall here. We compute the volumetric tensor in lag space $L_{jk} = \frac{1}{N}\sum_{\alpha=1}^N(\ell_{\alpha j}\ell_{\alpha k}-\ell_{b j}\ell_{b k} )$ where $\ell_{\alpha j}$ is the j-th component of the vertex $\alpha$, $\ell_{b j}=\langle\ell_{\alpha j}\rangle_\alpha$ is the j-th component of the mesocenter, and we use $N=4$. If we call $\lambda_1 \geq \lambda_2 \geq \lambda_3 $ the three eigenvalues of the volumetric tensor, we can finally define elongation (E) and planarity (P) parameters as $E=1-\sqrt{\lambda_2/\lambda_1}$ and $P=1-\sqrt{\lambda_3/\lambda_2}$ respectively. We will consider suitable for our analysis all tetrahedra that in the EP plane lie no further than 0.85 from the origin as shown in Fig.~\ref{fig:EP}. This rejects tetrahedra that are nearly planar or colinear as these would give unreasonable results to Eq.~\ref{eq:yaglom}. Employing this procedure, we generally obtain more than 100 cascade rate estimates from a single sample of MMS data.

\section{Results}

\begin{figure}[ht]
    \centering
    \includegraphics[width=0.7\columnwidth]{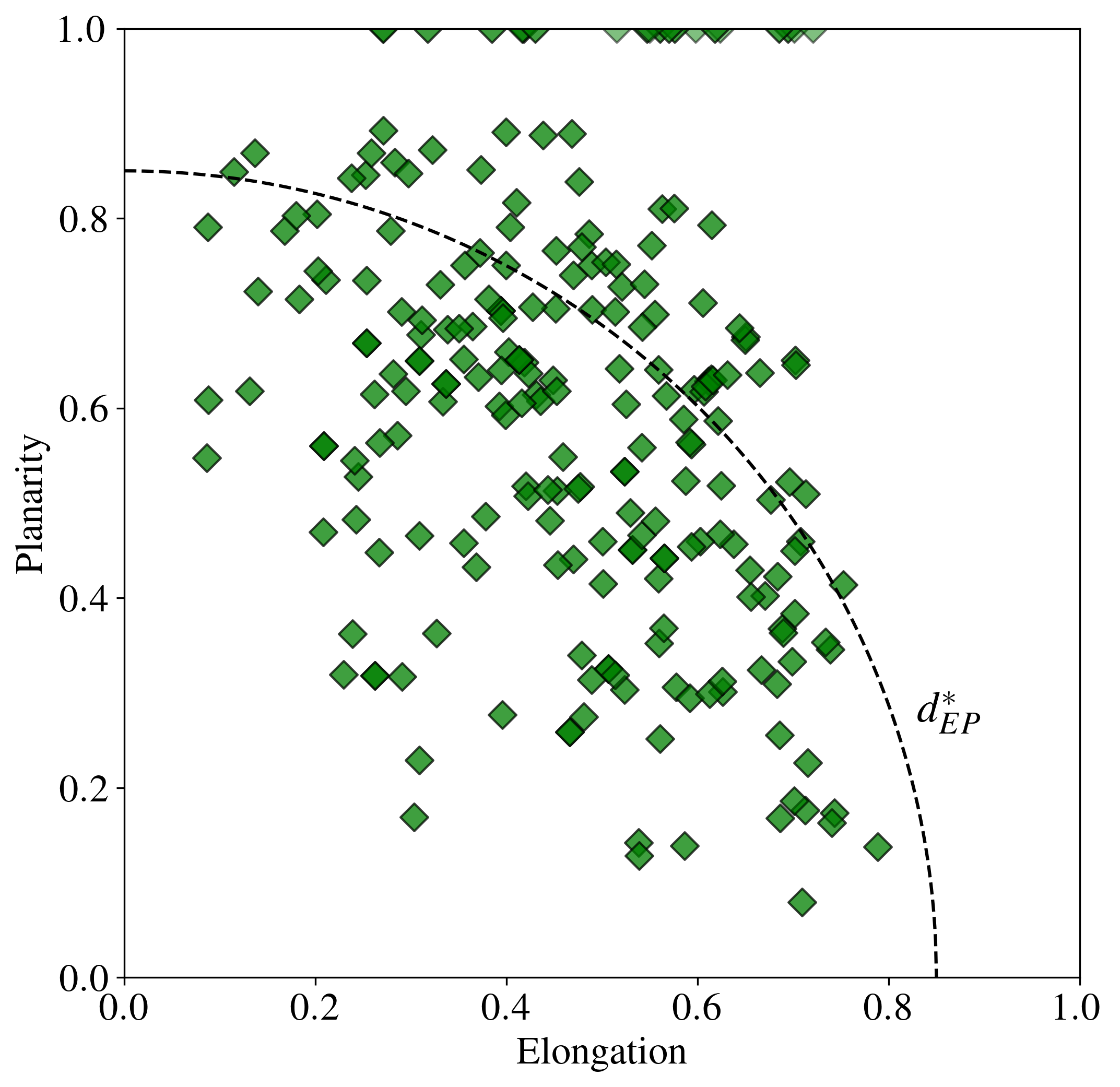}
    \caption{Quality check for tetrahedra. The dashed line indicates the threshold $d_{EP}^* = \sqrt{E^2 + P^2} = 0.85$. Tetrahedra with $d_{EP} \leq d_{EP}^*$ are used for further analysis, and the others are discarded. }
    \label{fig:EP}
\end{figure}

We compute the Yaglom flux $\YY$ using the 12 available baselines of MMS as described in the previous section. Finally, we evaluate the divergence by arranging the 12 points four at a time in all the possible 330 independent sets. This procedure has been applied to all intervals listed in Table~\ref{tab:table1}. Since the results are qualitatively similar, here we show those pertaining to interval IV (2017 Dec 26). For each tetrahedron, we compute the elongation and planarity parameters in order to later retain only those that are well-behaved that is, those that have a distance from the origin in the EP plane $d_{EP} = \sqrt{E^2 + P^2}\leq 0.85$ \citep{paschmann1998analysis}. We also verified that the result is not affected by this threshold. Even when selecting only those tetrahedra such that $d_{EP}\leq0.6$, we obtain the same value for the cascade rate. However, the number of estimates dropped from 134 to 27 and therefore we decided to keep the larger number of available points for statistical purposes. In Fig.~\ref{fig:EP}, we report the elongation and planarity values for the lag-space tetrahedra, together with the chosen threshold $d_{EP}^*=0.85$. Tetrahedra in lag space give a rather uniform coverage of the EP plane, with a few points at $P=1$ indicating flat geometries.

\begin{figure}[ht]
    \centering
    \includegraphics[width=0.9\columnwidth]{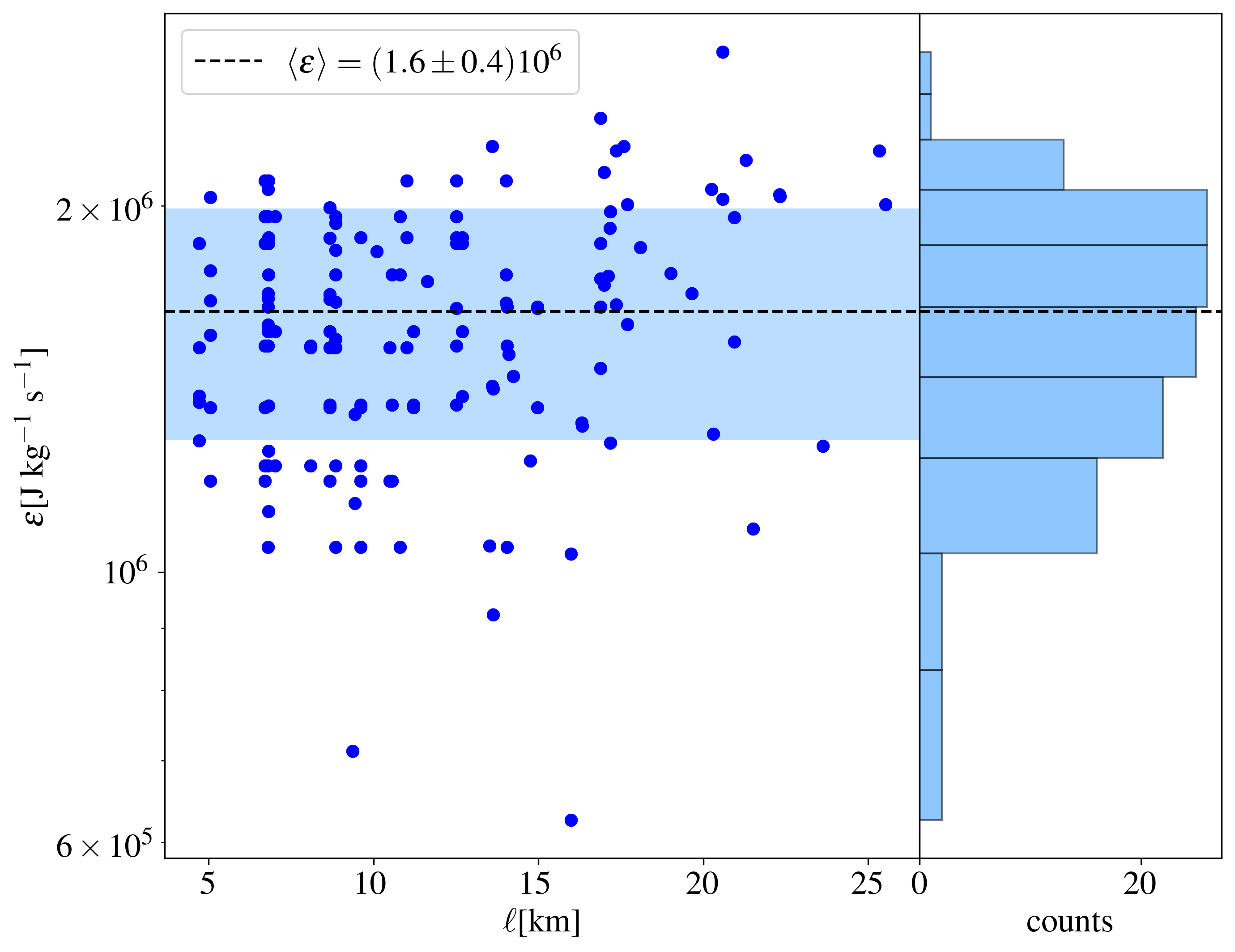}
    \caption{Scatter plot of the cascade rate as a function of lag. On the side, the histogram of the results is reported. The horizontal black dashed line indicates the average of the histogram, and the blue shaded region represents the uncertainty around the mean.}
    \label{fig:eps_ell}
\end{figure}

\begin{figure}[ht]
    \centering
    \includegraphics[width=0.85\columnwidth]{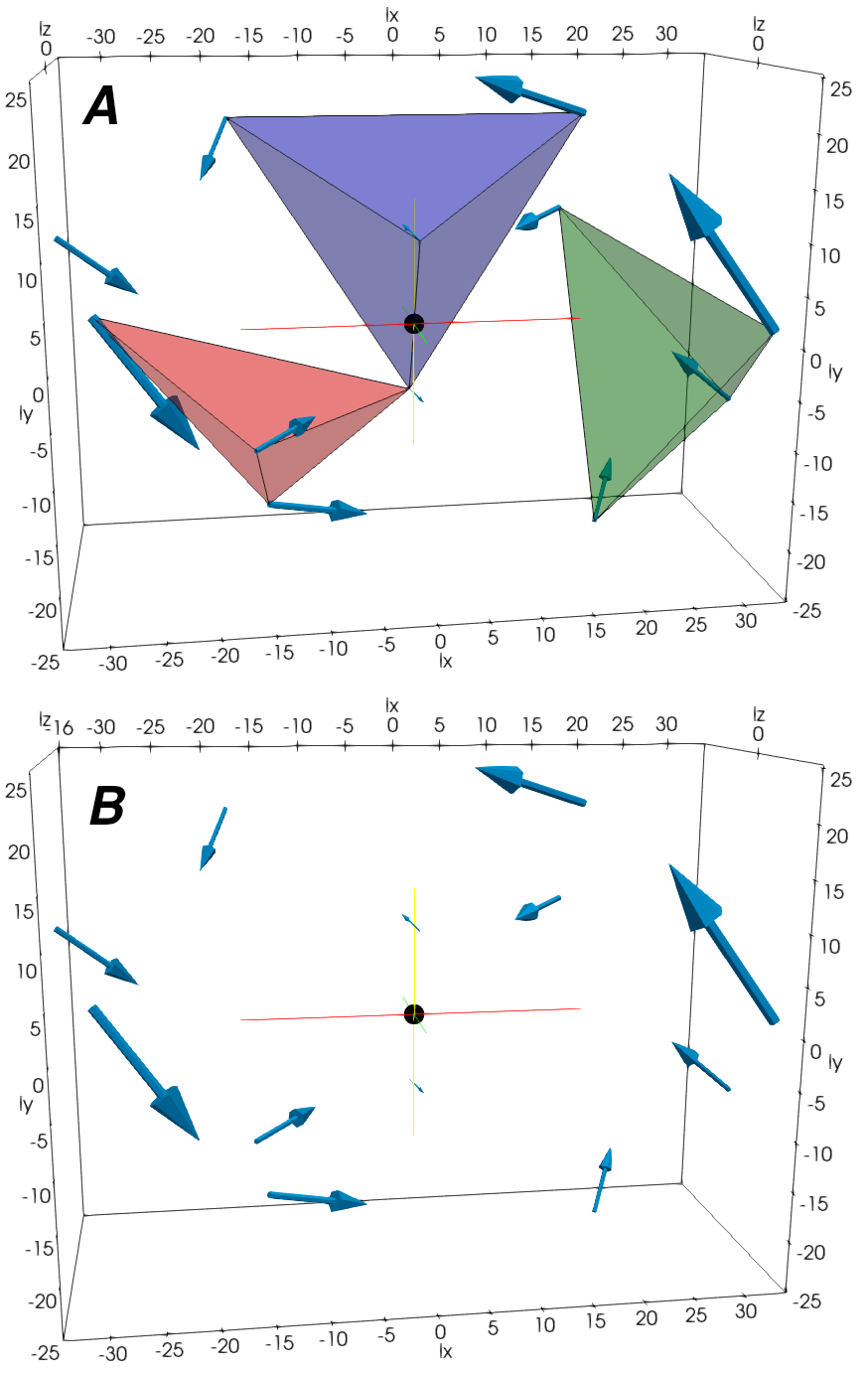}
    \caption{(A) Yaglom flux vectors in lag space with some tetrahedra, over which Yaglom's law is solved (shaded volumes). (B) same plot without highlighted tetrahedra. It is possible to appreciate the swirling motion of the Yaglom flux pointing toward the origin (indicated by the black sphere).}
    \label{fig:fluxarrow}
\end{figure}

In solving Yaglom's equation, each estimated  value of the cascade rate is assigned to the increment corresponding to the mesocenter of the tetrahedron that is used to compute the divergence. The scatter plot of the values of the cascade rate as a function of the position of the mesocenter in lag space is shown in Fig.~\ref{fig:eps_ell}. A visualization of some tetrahedra used to compute the divergence is given in Fig.~\ref{fig:fluxarrow}A.

From the average, the value that we obtain for the cascade rate $\langle \epsilon \rangle = ( 1.6 \pm 0.4 )\,10^6$ J kg$^{-1}$ s$^{-1}$. The width of the distribution suggests an uncertainty of $25\%$. The obtained values for the cascade rates are comparable to those present in literature \citep{bandyopadhyay2018incompressive}. However, they are more robust since we do not need to assume isotropy \citep{wang2022strategies}, and the large number of estimates per interval allows considering the signed value \citep{hadid2018compressible}. For estimation of uncertainties in simulations, see \citet{pecora2023helioswarm}. In general, the errors become much smaller when the number of available observation positions is increased, as will be the case in future multispacecraft missions.

The other quantity that might be of interest is the Yaglom flux itself $\YY$ as it could be imagined as the flux that nonlinearly transfers energy across scales. For a homogeneous, isotropic system, one expects the Yaglom flux to have magnitudes proportional to the distance from the origin, and radially pointing toward it.

Previous numerical studies \citep{verdini2015anisotropy} provided a picture of how the Yaglom flux behaves in lag space. In particular, it was confirmed that it is larger in magnitude going far away from the origin, and, when the system is anisotropic, it shows some deflections from the radial trend. In space plasmas, it has not been possible to visualize the behavior of the Yaglom flux prior to the development of this technique. 
From the present results, we know the flux vector at 6 (12) points in the lag space and can examine its behavior, as shown in Fig.~\ref{fig:fluxarrow}B. This visualization of the Yaglom flux using space data confirms the basic features that were expected. The arrows' lengths in the Figure are proportional to the magnitude of the Yaglom flux, and we notice that they become smaller as they approach the origin of the system where dissipation will eventually deal with the energy that has been transferred there. Moreover, we observe a swirling of the arrows around the origin instead of radially pointing toward it as it would if the turbulence was isotropic.

\section{Discussion and conclusions}
We have applied a novel multispacecraft technique that was developed by \citet{pecora2023helioswarm} to MMS measurements in the magnetosheath to evaluate the turbulence cascade rate due to the Yaglom term in the vkH equation. The technique is extended by exploiting the symmetry properties of the mixed third-order structure function. This provides a statistically significant $(>100)$ number of estimates of the cascade rate by computing the full divergence in Eq.~\ref{eq:yaglom}. 

Previously cascade rate estimation in space plasmas has been limited to one value per interval and mostly employing a 1D version of Yaglom's equation \citep[e.g.][]{sorrisovalvo2007observation}. That approximation, however, assumes isotropy of the system, as well as the usage of the Taylor hypothesis that convolves space and time correlations. This major advancement allows having a distribution of values for the (partial) cascade rate and, possibly, a value closer to the real one. Moreover, we are also able to visualize for the first time in space plasmas the Yaglom flux that is responsible for transferring energy toward smaller scales from the inertial range. We indeed see that the Yaglom flux magnitudes in the magnetosheath become smaller when approaching smaller and smaller scales where energy will eventually be dissipated. The swirling of the flux vectors, a departure from the often-assumed radial convergence is possibly due to the anisotropy of the system, uncertainty in the data, and possibly other unknown causes that need to be further investigated. 

This is the first milestone in the employment of multispacecraft techniques to measure the anisotropic cascade rate in space plasma. The current limitations such as restrictions to the magnetosheath and to scales smaller than the inertial range will be overcome once the next generation of multiscale multispacecraft missions such as HelioSwarm  \citep{spence2019helioswarm} and Plasma Observatory \citep{retino2022particle_PO} appear in the solar wind.

\section{Acknowledgments}
Supported by NASA MMS Mission under grant number 80NSSC19K0565 at the University of Delaware and 80NSSC21K0454, 80NSSC22K0688 grants at the University of Colorado Boulder.



\end{document}